\documentclass[a4paper,useAMS,usenatbib]{mnras}

\usepackage{graphicx}
\usepackage{setspace}
\usepackage{natbib}
\usepackage{color}
\usepackage{amsmath,amssymb}
\usepackage{times}
\usepackage{hyperref}

\AtBeginShipout{%
  \ifnum\value{page}>1 %
    \typeout{* Additional boxing of page `\thepage'}%
    \setbox\AtBeginShipoutBox=\hbox{\copy\AtBeginShipoutBox}%
  \fi
}

\newcommand{\Gyr}            {\,{\rm Gyr}}

\newcommand{\kpc}            {\,{\rm kpc}}
\newcommand{\Mpc}            {\,{\rm Mpc}}
\newcommand{\Msun}           {\,{\rm M}_\odot}
\newcommand{\kms}            {\,{\rm km}\,\,{\rm s}^{-1}}
\newcommand{\kmsMpc}         {\,\,{\rm km}\,\,{\rm s}^{-1}\,{\rm Mpc}^{-1}}
\newcommand{\feh}            {{\rm [Fe/H]}}
\newcommand{\vth}            {V_{\rm \theta}} 
\newcommand{\vr}             {V_{\rm r}}

\newcommand{\fig}      {Fig.~\ref}
\newcommand{\bc}       {\begin{center}}
\newcommand{\ec}       {\end{center}}

\newcommand {\radial}     {radially-biased }

\title[Gaia Sausage-like components in Auriga]{The origin of galactic metal-rich stellar halo components with highly eccentric orbits}

\author[A. Fattahi et al.]{Azadeh Fattahi$^{1}$\thanks{E-mail: azadeh.fattahi-savadjani@durham.ac.uk},
Vasily Belokurov$^{2,3}$,
Alis J. Deason$^{1}$,
Carlos S. Frenk$^{1}$, \newauthor 
Facundo A. G{\'o}mez$^{4,5}$,
Robert J. J. Grand$^{6,7,8}$, 	
Federico Marinacci$^{9}$, \newauthor
R{\"u}diger Pakmor$^{8}$, 
Volker Springel$^{8}$
\\
$^{1}$Institute for Computational Cosmology, Department of Physics, University of Durham, South Road, Durham DH1 3LE, UK\\
$^{2}$Institute of Astronomy, Madingley Road, Cambridge CB3 0HA \\
$^{3}$Center for Computational Astrophysics, Flatiron Institute, 162 5th Avenue, New York, NY 10010, USA \\
$^{4}$Instituto de Investigaci{\'o}n Multidisciplinar en Ciencia y Tecnolog{\'i}a, Universidad de La Serena, Ra{\'u}l Bitr{\'a}n 1305, La Serena, Chile \\ 
$^{5}$ Departamento de F{\'i}sica y Astronom{\'i}a, Universidad de La Serena, Av. Juan Cisternas 1200 N, La Serena, Chile \\ 
$^{6}$ Heidelberger Institut f{\"u}r Theoretische Studien, Schlo{\ss}-Wolfsbrunnenweg 35, 69118 Heidelberg, Germany \\ 
$^{7}$ Zentrum f{\"u}r Astronomie der Universit{\"a}t Heidelberg, Astronomisches Recheninstitut, M{\"o}nchhofstr. 12-14, 69120 Heidelberg, Germany \\ 
$^{8}$ Max-Planck-Institut f{\"u}r Astrophysik, Karl-Schwarzschild-Str. 1, D-85748, Garching, Germany\\
$^{9}$ Harvard-Smithsonian Center for Astrophysics, 60 Garden St., Cambridge, MA 02138, USA 
}

\date{Accepted XXX. Received YYY; in original form ZZZ}

\pubyear{2015}

\begin{document}
\label{firstpage}
\pagerange{\pageref{firstpage}--\pageref{lastpage}}
\maketitle

\begin{abstract}
  Using the astrometry from the ESA's \textit{Gaia} mission, previous works
  have shown that the Milky Way stellar halo is dominated by 
  metal-rich stars on highly eccentric orbits. 
  To shed light on the nature of this prominent halo
  component, we have analysed 28 Galaxy analogues in the Auriga suite of
  cosmological hydrodynamics zoom-in simulations. Some three quarters of the Auriga
  galaxies contain prominent components with high radial velocity
  anisotropy, $\beta > 0.6$. However, only in one third of the hosts 
  do the high-$\beta$ stars contribute significantly to the accreted stellar
  halo overall, similar to what is observed in the Milky Way. For this 
  particular subset we reveal the origin of the dominant stellar halo component
  with high metallicity, [Fe/H]$\sim-1$, and high orbital anisotropy, $\beta>0.8$, 
  by tracing their stars back to the epoch of accretion. 
  It appears that, typically, these stars come from a single
  dwarf galaxy with a stellar mass of order of $10^9-10^{10}\Msun$ that merged 
  around $6-10 \Gyr$ ago, causing a sharp increase in the halo mass. 
  Our study therefore establishes a firm link between the
  excess of radially anisotropic stellar debris in the halo and an
  ancient head-on collision between the young Milky Way and a massive
  dwarf galaxy.
  
\end{abstract}

\begin{keywords}
galaxies: Milky Way, galaxies: dynamics
\end{keywords}



\section{Introduction}

The history of mass accretion onto the Milky Way can be constructed from
the study of unrelaxed tidal debris in its stellar halo
\citep[e.g.][]{DeLucia2008,Johnston2008,cooper10}. This works particularly well
for the most recent events, such as the infall of the Sagittarius (Sgr)
dwarf \citep[see e.g.][]{Ibata1994,Majewski2003}. Yet, despite the fact that
the Sgr stream has been mapped in glorious detail over the last decade
or so
\citep[see][]{Ibata2001,Majewski2003,Newberg2003,Belokurov2006,Yanny2009,Koposov2012},
we have been remarkably slow to comprehend the damage this dwarf
galaxy has been inflicting onto the Milky Way
\citep[see][]{Purcell2011,Gomez2013,Laporte2018}. Sgr is a striking example of
how much destruction a massive satellite can bring to the fragile
Milky Way's disc even on an orbit of moderate eccentricity
\citep[e.g.][]{Price-Whelan2015,Hayes2018,Bergemann2018,deBoer2018,Deason2018}.

Events in the distant past are normally less obvious, blurred by the
debris' gradual relaxation in the host's gravitational potential and
concealed behind the layers of the subsequent Galactic growth
\citep[e.g.][]{Jorge2006,Gomez2010,Buist2015,Erkal2016}. Nonetheless,
some episodes in the Milky Way's history are so cataclysmic that they
are difficult to hide, even if they took place a long time ago.

It has been known for some time that the properties of the Milky Way stellar
halo change drastically as a function of Galactocentric radius. This
is revealed in the rapid variation of the stellar halo's kinematics
and metallicity with radius \citep[see e.g.][]{Chiba2001,Carollo2007,Carollo2010}
as well as the halo's density profile
\citep[see][]{Watkins2009,Deason2011,Sesar2011}. Two possible
explanations have been put forward, one invoking the presence of the
so-called in-situ halo
\citep[][]{Zolotov2009,McCarthy2012,Tissera2013,Cooper2015} and one appealing to
an early accretion of a massive galaxy \citep[][]{Deason2013}. While
unambiguous evidence for an in-situ population is yet to be found
\citep[see e.g.][]{Haywood2018}, the hypothesis of a single accretion
event dominating the nearby Galactic halo has been reinforced by
multiple independent observations. These include the pattern of the
stellar halo's alpha-abundances
\citep[see][]{Venn2004,Tolstoy2009,Nissen2010,deBoer2014,Hayes2018b}, the ratio of Blue
Straggler to Blue Horizontal Branch stars \citep[][]{Deason2015}, the
evolution in the mixture of RR Lyrae of different Oosterhoff types
\citep[][]{unmixing} and the relative scarcity of the Galactic stellar
halo substructure \citep[][]{Helmi2011,Xue2011,smoothness}.

Most recently, with the help of the astrometry from the ESA's
\textit{Gaia} mission \citep[][]{Prusti2016}, the Milky Way halo has
been demonstrated to contain a large number of metal-rich stars on
extremely eccentric orbits
\citep[][]{Belokurov2018,ActionSpace,Haywood2018,Helmi2018,Mackereth2018}. Using the established
mass-metallicity relationship \citep[][]{Kirby2013} as well as
numerical simulations of Galaxy formation, \citet{Belokurov2018} have
argued that the observed chemistry and the drastic radial anisotropy
of the halo stars (the ``\textit{Gaia} Sausage'') are telltale signs of an
ancient collision between the Milky Way and a massive dwarf. This
conjecture builds on the insights into the stellar halo assembly
presented in the work by \citet{Amorisco2017}, which for the first
time posits a link between high radial anisotropy and the infall of
a massive satellite.  The idea of an ancient intergalactic crash has
since received additional support with the discovery of a large number
of globular clusters associated with this accretion event
\citep[see][]{SausageClusters, Kruijssen2018}. Additionally, the
enormous extent of the ``Sausage'' debris is now starting to become
apparent with the studies of the stellar halo's anisotropy variation
with radius \citep[][]{Bird2018,Lancaster2018}. The break in the
stellar halo density has been revealed to correspond to the apocentric
pile-up of the ``Sausage'' stars \citep[see][]{pileup}, while the
large diffuse halo overdensities such as Hercules-Aquila and Virgo
Clouds are now believed to be linked to the same accretion event
\citep[see][]{Simion2018}. While most of these studies rely on
small subsamples of halo tracers with complete phase-space
information, \citet{Wegg2018} and \citet{Iorio2018} have recently used
large, all-sky catalogs of RR Lyrae with proper motions to decipher
the kinematic properties of the halo. As these authors demonstrate, the
bulk of the Galactic stellar halo between 5 and 30 kpc - as traced by
these old pulsating stars - is on strongly radial orbits, yielding
estimates of $\beta \sim 0.9$.

The dominance of one massive accretion event is in good agreement with
the general predictions of cosmological simulations. Several works
have shown that the majority of stellar mass accreted onto Milky Way
mass haloes is contributed by a small number of massive satellites
\citep[e.g.][]{bullock05, cooper10, deason16}. The hypothesis of a
small number of massive contributors is a direct consequence of the
hierarchical dark matter model, and the extremely steep
stellar-to-halo mass relation in the dwarf galaxy regime 
\citep[e.g.][]{Eke2004,Moster2013,gk14,brook14, 
Sawala2016a,read17,Jethwa2018,Fattahi2018,Simpson2018}. Thus, despite the
large numbers of low-mass dwarfs accreted by Milky Way-mass galaxies,
the contribution of these low mass dwarfs to the \textit{total} accreted stellar 
mass is small \citep[see][]{deason16}. 

Curiously, some of the first efforts to study the emergence of orbital
patterns in simulated stellar haloes already saw trends similar
to those recently discovered in the Milky Way \citep[see
  e.g.][]{Brook2003,Meza2005,Navarro2011}. While those works were limited to a
small number of hand-picked cases, here we strive to carry
out a systematic analysis using a statistically significant sample of
realistic simulations of Milky Way analogues. \cite{monachesi18} recently studied the
global properties of stellar haloes in the Auriga simulations \citep{Grand2017}. Indeed,
in agreement with earlier work, these authors find that the global
properties of the stellar haloes, such as the average metallicity,
total stellar halo mass, and density profile, are directly related to
the small number of stellar halo progenitors.  

In this work we use the
Auriga simulations \citep{Grand2017} to understand the relatively local stellar halo
kinematics, within 20 kpc of the Galactic centre\footnote{Note that
  \cite{monachesi18} studied the stellar halo components at larger
  distances of $10-100 \kpc$.}, to compare with the recent \textit{Gaia} results. Ours
is not the first attempt to interpret the observed local stellar
halo properties with the help of simulated Milky~Way 
like haloes. Most recently, \citet{Mackereth2018} compared \textit{Gaia} and 
APOGEE data with the orbital and chemical properties of the accreted 
halo stars in the EAGLE simulation suite \citep[see][]{Schaye2015,Crain2015}. They
come to the conclusion that to explain the measured $\alpha$ vs [Fe/H]
trend as a function of eccentricity, an early accretion of a massive
dwarf galaxy is required. A similar conclusion is reached by
\citet{Emma2018} who use the stellar haloes in hydrodynamical 
simulations of Aquarius MW-like haloes \citep[Aquila, see
  e.g.][]{Tissera2013,Scannapieco2009,Springel2008b}.
  The Auriga simulations used here provide a higher resolution 
  compared to EAGLE, and it gives us access to a significantly
larger sample size with richer physics compared to Aquila. 

Our Paper is structured as follows. In Section 2 we describe the
\textit{Gaia} Data Release 2 observations of a ``Sausage-like"
component in the phase space of the stellar halo. We describe the
Auriga simulations in Section 3 and make a comparison with the
\textit{Gaia} results. In particular, we show how these highly
eccentric, metal-rich stellar halo components (referred to as the
\radial components, hereafter) are also produced in the
simulations. Finally, in Section 4 we summarise and discuss our main
results.

\section{Observations}
\label{sec:obs}
\begin{figure}
  \centering
  \includegraphics[width=0.45\textwidth]{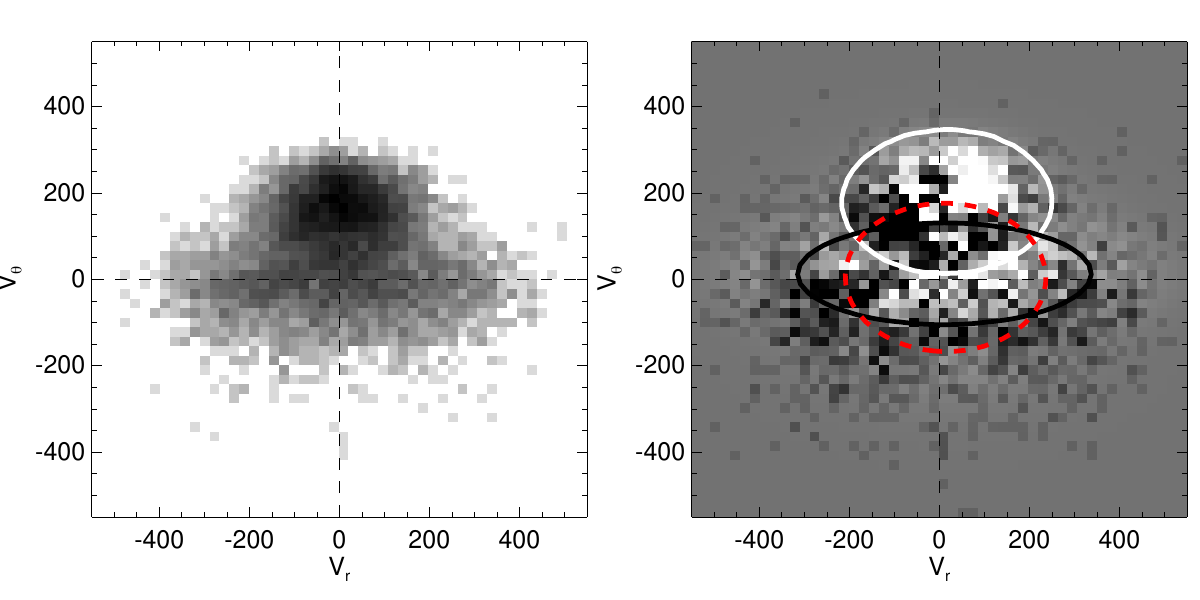}
  \caption[]{The ``\textit{Gaia} Sausage'' as viewed with the GDR2 sample. {\it
      Left:} Logarithm of the stellar density in spherical $v_r,
    v_{\theta}$ coordinates, corresponding to the radial and azimuthal
    velocity components. Two prominent overdensities are visible, one
    corresponding to the (thick) disc with $\bar{v}_{\theta}\sim170
    \kms$ and one corresponding to the halo with no apparent
    rotation. {\it Right:} Residuals of the multi-variate three
    component Gaussian mixture model. Here excess (depletion) of the
    data with respect to the model is shown as a dark (light)
    region. Two thirds of the halo stars are in a radially-biased
    ($\beta\sim0.86$) ``Sausage'' component indicated with a solid black
    line. The rest of the halo is in a much more isotropic component
    with $\beta\sim0.4$ shown as dashed red line. Finally, at least
    half of this local GDR2 sample belongs to the disc, marked with a
    white solid line. }
   \label{fig:sausage_dr2}
\end{figure}

As a point of reference we use the view of the Milky Way's stellar
halo provided in Data Release 2 \citep[GDR2, see][]{Brown2018}
of the \textit{Gaia} mission \citep[see][]{Prusti2016}. Specifically, we use a
sample of GDR2 stars with radial velocities measured by \textit{Gaia}
\citep[see][]{Cropper2018,Sartoretti2018,Katz2018}. Out of
$\sim7\times10^6$ stars with available spectroscopic information, we
select $\sim12,000$ with good parallaxes,
$\varpi/\sigma_{\varpi}>3$ and $2<|z|/\kpc<4$,
$|b|>25^{\circ}$ and [Fe/H] $\leq-1$. The latter three of the above
cuts, including that on the \textit{Gaia}'s template metallicity
\citep[see][for details]{Katz2018}, are applied to
minimise the Galactic disc contamination. We convert the celestial
positions and parallaxes into Galactocentric Cartesian coordinates,
assuming a Solar radius of 8 kpc and a Sun's height above the
Galactic plane of 0. The proper motions and the line-of-sight
velocities are translated into velocity components in spherical
polars, using the Solar peculiar motion from \citet{LSR} and
marginalising over the Local Standard of Rest (LSR) values assuming 
they are Gaussian distributed\footnote{We are neglecting the asymmetric drift 
here, as we are interested in the $\vth \sim 0$ component where this effect is small.} 
with a mean at $238 \kms$ and a standard
deviation of $9 \kms$ \citep[see][]{Ralph2012}. The uncertainties are
propagated using Monte-Carlo sampling and the medians of the
 distributions for each object, as well as the scaled median absolute
deviations, are used for the analysis below.

Fig.~\ref{fig:sausage_dr2} shows the distribution of the median
radial, $v_r$, and azimuthal, $v_{\theta}$, velocity components in the
sample described above. Two populations are immediately apparent: 
the Galactic disc, with significant mean rotation and the
non-rotating Galactic halo. We estimate the properties of each
component by modelling the 3-D velocity distribution with a mixture of
three multivariate Gaussians using the Extreme Deconvolution code
\citep[ED, see][]{ED} which takes the measurement uncertainties into
account. In this model, the Gaussians represent the disc
and the two components of the halo, a radially-biased and a more
isotropic one. The results of the ED fitting exercise are shown in the
right panel of Fig.~\ref{fig:sausage_dr2}, where the residuals of
data minus model are shown as a greyscale, with dark
(white) regions corresponding to an excess (lack) of data points
compared to the model prediction. Overlaid are the isodensity
contours of the three Gaussians given by the best-fit model, with
white solid representing the disc, black solid the radially-biased
halo, and red dashed the second, more isotropic
halo component.

As inferred from the ED modelling, 50\% of our sample belongs to the
(thick) disc with a mean rotation of $\bar{v}_{\theta}\sim170 \kms$,
37\% belong to the radially-biased halo population with $\beta\sim
0.86$ and 13\% to the halo component with $\beta\sim 0.4$. More
specifically, the radially-anisotropic ``Sausage'' component, whose
radial, azimuthal and polar velocity dispersions are 177, 61 and 70
$\kms$ respectively, contains just over two thirds of the total halo
population locally in good agreement with
\citet{Belokurov2018}. Interestingly, according to this Gaussian
mixture model, the \radial component is centered on zero in all three
velocity components, i.e. it exhibits no net rotation, in mild tension
with some of the earlier results
\citep[e.g.][]{Belokurov2018,Necib2018}, where a prograde rotation of
$\sim20 \kms$ was reported. The quality of the fit can be judged from
the right panel of Fig.~\ref{fig:sausage_dr2}. Leaving aside the
residuals at high $v_{\theta}$ associated with the disc, there is a
dark/light pattern of under/over subtraction of the model from the
data. The excess of observed stars with high $|v_r|$ velocities is
consistent with previous modelling attempts. However, a less prominent
but noticeable ridge of positive residuals exists at negative
$v_{\theta}$. This implies that there may be a significant spray of
retrograde debris possibly associated with the ``Sausage''
\citep[see][]{ActionSpace, Helmi2018} not accounted for by our
model. Judging by the shape of the data in the left panel of the
figure as well as the residuals in the right, we conjecture that these
residuals are likely due to the strongly non-Gaussian shape of the
velocity distribution in the ``Sausage'' component. The complicated
structure of the residuals advocates the use of more sophisticated
models of the ``Sausage'' kinematics as those recently proposed in
\citet{Necib2018} and \citet{Lancaster2018}. We, however, choose to 
use the relatively simple ED modeling here that can 
be applied to the simulations as well. This will enable us 
to make a comparison between the simulations and observations, 
and also between different simulated galaxies.

\section{Auriga simulations}

We use cosmological magneto-hydrodynamical (MHD) simulations of MW-like haloes from the Auriga project \citep{Grand2017} to interprete the origin of the ``\textit{Gaia} Sausage" described in the previous section. The Auriga project includes a suite of 30 MW size haloes simulated at high resolution, using the ``zoom-in'' technique \citep{Frenk1996,Jenkins2013} and the Tree-PM moving-mesh code {\small Arepo} \citep{Springel2010}. The simulations have been performed at different levels of resolution. We use the standard resolution runs (level-4) in this study rather than the higher resolution ones (level-3), as only eight haloes have been run at level-3. The mass resolution of the standard level-4 runs is  $5\times10^4 \Msun$ and $3\times10^5 \Msun$ per gas and dark matter particles, respectively. The selection of MW candidates, as well as the subgrid physics model, have been described in detail in \citet{Grand2017}. Here we give a brief summary.

MW-like haloes were selected from the $100^3 \Mpc^3$ dark matter-only periodic box of the {\small EAGLE} project \citep{Schaye2015,Crain2015}, based on their virial\footnote{Virial quantities are defined by reference to a mean overdensity of 200 times the critical density.} mass, $\sim 10^{12}\Msun$, and an isolation criterion. The subgrid physics model in the re-simulations includes a spatially uniform photoionizing UV background, primordial and metal line cooling, star formation, stellar evolution and supernovae feedback, supermassive black hole growth and feedback, and magnetic fields. A fixed set of chemical elements, including Fe and H, have been consistently tracked within the simulations.    
Haloes and bound structures and substructures were identified using the FoF and {\small SUBFIND} algorithms \citep{Davis1985,Springel2001a}, respectively. We refer to the central galaxy of the main halo in each Auriga simulation as a MW analogue. The cosmological parameters adopted for the simulations are those of Planck Collaboration XVI \citep{Planck2014}: matter density of $\Omega_m=0.307$, baryonic density of $\Omega_b=0.048$, and a Hubble constant of $H_0 = 67.77 \kmsMpc$.

We use all 30 MW analogues of the level-4 (standard) runs, except for Au-11 and Au-20. These two galaxies are undergoing a merger at the present time and this distorts the phase space distribution of star particles around the galactic discs.

The coordinate and velocity reference frame of the MW analogues are based on {\small SUBFIND}. The galactic discs are defined, and visually confirmed, using the net angular momentum of star particles in the inner $10\kpc$. For our stellar samples, we keep star particles between $r=5-20\kpc$ from the galaxy centre, with galactic latitude $|b|>15 \deg$, assuming a random azimuthal angle for the position of the Sun at 8kpc to roughly mimic the observational selections of \citet{Belokurov2018} \footnote{We have checked that the results are not sensitive to these cuts.}. We only include stars bound to the MW analogues (not to the satellites) and shift the $\feh$ of star particles by a constant value of $\sim 0.5$ to roughly match the median and the [5th,95th] percentile of an equivalent sample in SDSS \citep{ahn12}. We adopt -2.0, -0.83, and -0.23 as the 5th, 50th, and 95th percentiles for SDSS stars, respectively.

We identify ``accreted'' star particles as those that were bound to galaxies other than the 
main progenitors of the MW analogues, at the snapshot following their formation time. 
According to this definition, stars that were formed out of the stripped 
gas from the satellites will be flagged as ``in-situ''
if they are bound to the MW analogues' progenitors right after their birth. 
We note that this identification method is dependent on the time 
resolution of the output snapshots. The temporal resolution is 
always better than $200$ Myr in Auriga level-4 simulations, and therefore, for our purposes, is not a major concern.      

\begin{figure*}
	\resizebox{15cm}{!}{\includegraphics{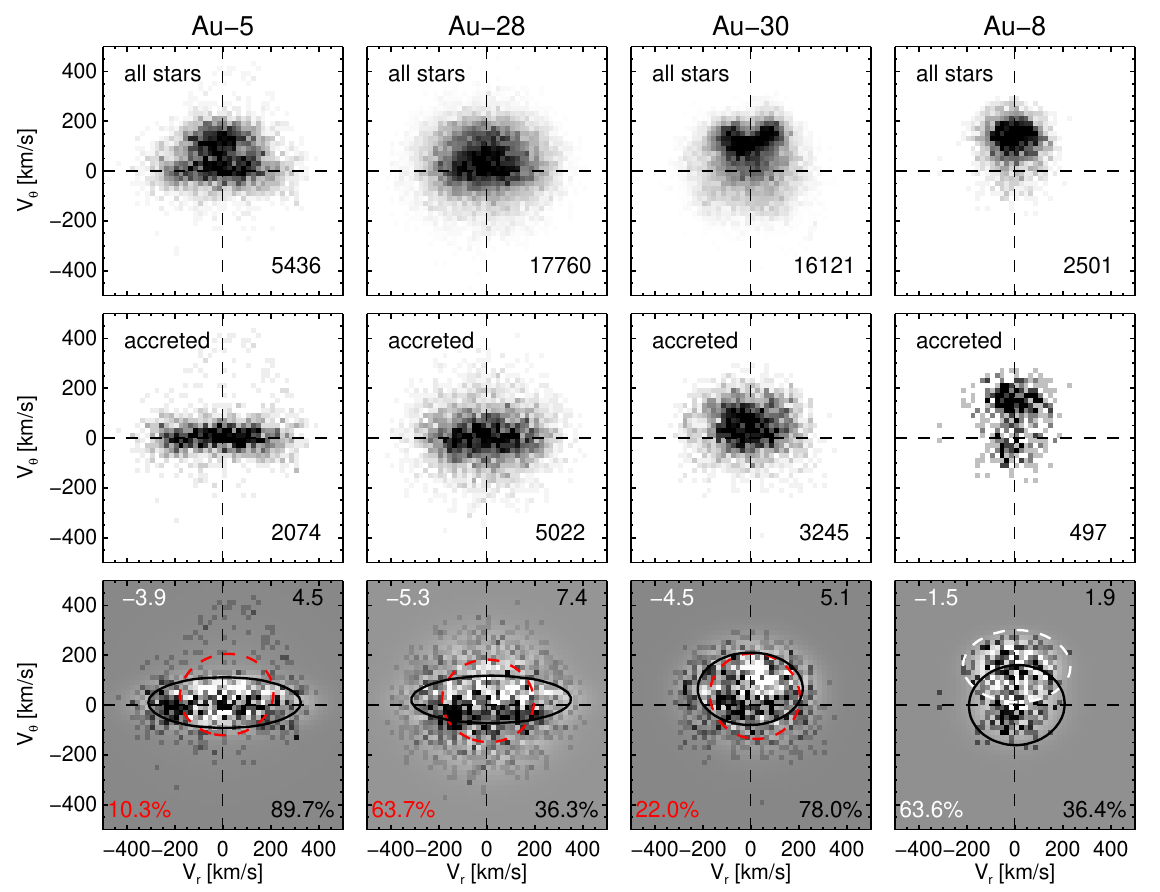}}
    \caption{Distribution of star particles in velocity space for four
      Auriga halos (Au-5, Au-8, Au-28, Au-30) shown in different columns. 
      The particles are located between
      $9-15\kpc$ above and below the disc and have $-1<\feh<-0.7$.
      These examples illustrate the variety of kinematic features
      found in the inner stellar halo in Auriga galaxies. Au-5 and Au-28 have a
      highly radial population of stars, while Au-8 and Au-30 lack such 
      a component. {\it Top:} Velocity space of all star particles in the 
      given height and metallicity bin. 
      The number of star particles in each panel is quoted in the legend.
      {\it Middle:} Similar to the top row but only for accreted 
      stars. {\it Bottom:} The residuals of the fits to
      the velocity space of the accreted stars. The fits include
      two components. The black (white) pixels represent
      excess (depletion) of data relative to the model. 
      The numbers in the top corners of
      each panel give the highest and lowest values of the residuals,
      while the percentages on the lower corners give the contribution
      of different components with corresponding colors.}
    \label{fig:examples}
\end{figure*}

\begin{figure}
	\resizebox{\columnwidth}{!}{\includegraphics{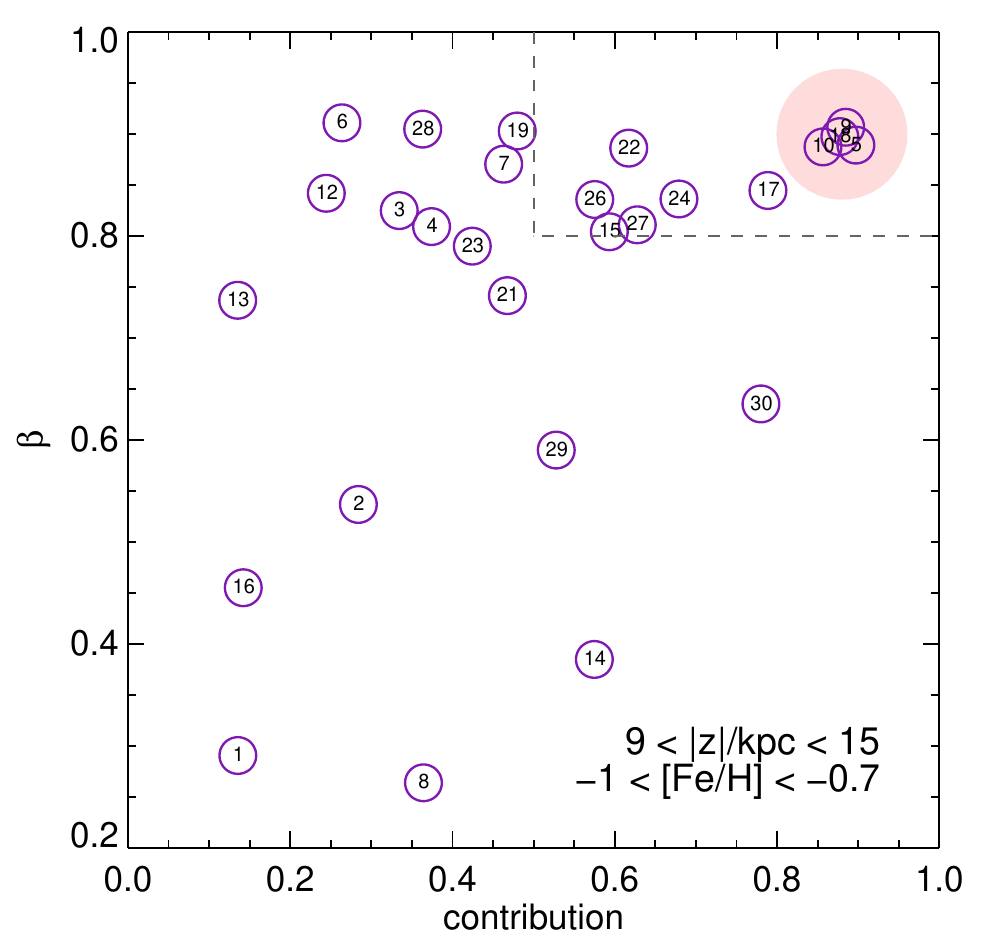}}
    \caption{Anisotropy parameter ($\beta$) of the more radial component of 
    	the Auriga stellar haloes, versus their relative contribution to the stellar halo. 
        These results are based on the decomposition of the velocity space (see 
        bottom panels of \fig{fig:examples} and the text for further details). We only consider 
        accreted stars at galactic height of $9<|z|/\kpc<15$ with $-1<\feh<-0.7$. 
        Galaxies inside the dashed lines have a dominant,
        highly radial component similar to the \textit{Gaia} Sausage. 
        The four haloes highlighted in the top-right corner (Au-5, 
        Au-9, Au-10, Au-18) are the 
        extreme cases, also marked in the next plots, where 
        a highly-radial component dominates their inner stellar halo. 
	   	The small numbers inside each symbol represent the
        halo number of the Auriga galaxies for reference.}
    \label{fig:beta}
\end{figure}

\subsection{The \radial stellar halo components in Auriga}

Following \cite{Belokurov2018}, we plot the distribution of stars
in a 2D velocity space: azimuthal ($\vth$) vs radial ($\vr$) in
spherical coordinates, at different galactic height bins and
metallicities for all our MW analogues. Examples of the velocity space
in a grid of height and $\feh$ are presented in Appendix~\ref{app:example}. Here we
focus our analysis on star particles at $9 < |z|/\kpc < 15$ and $-1 < \feh < -0.7$. 
This galactic height range is larger than what was used in
the MW analysis in Sec.~\ref{sec:obs}. The reason behind our choice is that the
differences between Auriga haloes are clearer at these heights and
metallicities. At lower heights (but similar $\feh$), the distribution
is dominated by stars in the galactic disc and at lower $\feh$ values,
the radially-biased components become weaker. Our choice of sample
selection criteria is reasonable since our goal is not to find an
exact match to the MW’s ``\textit{Gaia} Sausage” in the simulations, but rather
more generally to investigate the origin of components with similar
properties to those of the structure found in the MW.

The top row of \fig{fig:examples} shows the velocity space of four example haloes (Au-5, Au-8, Au-28, and Au-30), for star particles at $9<|z|/\kpc<15$ and $-1<\feh<-0.7$.
The rotating component with $\vth\sim200\kms$ in the top panels of \fig{fig:examples} corresponds to the disc\footnote{This component is weak in Au-28 because the disc is more prominent only at lower heights.}. The halo components of these examples, which have much lower rotation (i.e. $\vth \sim 0$), appear very different. Au-5 has a prominent \radial component, similar to the real MW in \fig{fig:sausage_dr2}; while Au-8 lack such a component and are much more isotropic. These two galaxies illustrate the two ends of the spectrum of inner stellar halo kinematic features in the Auriga MW analogues.

The second row of \fig{fig:examples} shows the velocity space of the accreted component only, with the same selection cuts as the top panels. \textcolor{black}{We note that a large number of stars are removed from the top row when we select only accreted stars. For the particular spatial and metallicity selections used in this figure, the fraction of accreted stars is on average 30 per cent amongst 28 Auriga haloes. The fraction increases to 40 per cent if we relax the metallicity cut, mainly because the stellar halo is more metal poor. These numbers are overall consistent with the results of \citet{monachesi18}. An exact comparison is not possible since they used different spatial and kinematic selections.}

\textcolor{black}{According to \fig{fig:examples}, the rotating (disc) component is, in most cases, not present in the accreted stars.} A \radial component stands out in both Au-5 and Au-28, while the latter case is less prominent. Au-30 and Au-8, on the other hand, are isotropic. Moreover, Au-8 has an accreted component in its disc. The existence of accreted (or ``ex-situ'' ) disc components is not surprising. \citet{Gomez2017} extensively discussed their origin in Auriga galaxies, and \citet{RodriguezGomez2016} also report their existence in the {\small Illustris} simulations. \fig{fig:examples} illustrates that the \radial component is more easily identified in the accreted stars. Moreover, the existence of an in-situ halo component in the MW is still under debate \citep[see, e.g.,][]{Helmi2011, Bonaca2017, Deason2017, Haywood2018}. We therefore include only accreted particles for the rest of our analysis, unless otherwise stated.

In order to quantify the kinematic components in Auriga we decompose the velocity space of stars, at different heights and metallicities, into multi-variate Gaussian components using the Extreme Deconvolution method, similar to Section \ref{sec:obs}. The details of the procedure are described in Appendix~\ref{app:fit}. In summary, we first combine star particles for each MW analogue in various $\feh$ bins, and then decompose the velocity distribution at each galactic height into either two or three Gaussians, depending on the existence of a rotating (accreted) disc component. The velocity ellipsoids have 6 free parameters describing their centre and shape, at this stage. The three component fits include one for the disc and two for the halo-like stars; the two component fits typically include two halo-like ones. We visually inspect all 28 Auriga simulations to check that the fits are reasonable. For the second stage of the fitting, we fix the centre and shape of the ellipsoids from the previous step, and fit them to different $\feh$ bins at each height, letting the amplitude (i.e. contribution) of the components vary. 



The bottom row of \fig{fig:examples} shows the residuals of the fits to the accreted stars. The ellipses illustrate the components of the fits: a \radial halo component (black ellipse) and a more isotropic one (red dashed ellipse) for Au-5 and Au-28; two relatively isotropic halo components for Au-30 where one (black ellipse) is slightly more radial than the other (red ellipse); an isotropic halo component (black ellipse) and a disc component for Au-8 \footnote{Au-8 is an exception where the two components of the fit include one for the disc and one for the halo. This is equivalent to a three component fit where the contribution of a \radial halo component is zero.}. The black ellipse of Au-5, which is equivalent to the \radial component observed in \textit{Gaia}, indicates $\beta\sim0.9$ with a large contribution to the stellar distribution ($\sim90$ per cent). The highly eccentric component of Au-28, however, contributes a much lower fraction. The contributions of each component are given by the percentages written in the bottom corners. 
We find that the \radial components, if they exist, become less prominent with decreasing metallicity (see \fig{fig:Au5} and \fig{fig:Au9fit}); similar to pattern in the observations of \cite{Belokurov2018}.

\begin{figure}
	\resizebox{\columnwidth}{!}{\includegraphics{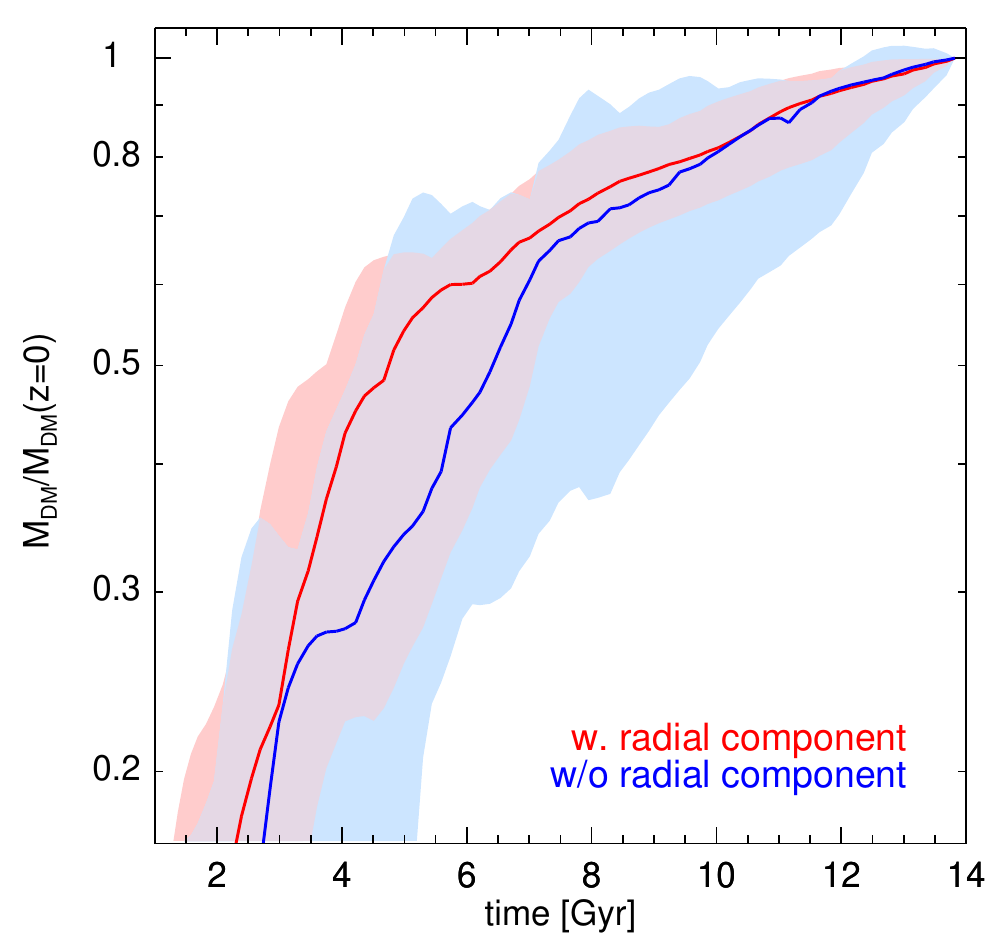}}
    \caption{Dark matter accretion history of Auriga galaxies, normalized to the 
    	present-day halo virial mass. The red line (shaded region) illustrates the median 
        (full range) of the accretion histories for galaxies with a significant 
        \radial stellar halo component (highlighted in the top-right corner of 
        \fig{fig:beta}). The blue line and shaded region correspond to the 
        accretion history of the other galaxies.}
    \label{fig:accretion}
\end{figure}

\fig{fig:beta} presents the properties of the {\it higher $\beta$
  component} of the stellar haloes as inferred by ED, at
$9<|z|/\kpc<15$ and $-1<\feh<-0.7$, in the 28 Auriga galaxies. A
significant fraction of these galaxies, $\sim75\%$, has a halo
component with high anisotropy ($\beta>0.7$). A key point to consider
is the component's contribution to the (accreted) stellar 
halo within the spatial and metallicity selection cuts 
(see examples in \fig{fig:examples}). 
Recall that the \radial component of the MW 
dominates the mass of the inner stellar halo, as discussed in
Sec.~\ref{sec:obs}. Therefore, we identify haloes with high $\beta$
{\it as well as} a large fractional contribution to the halo (the top-right corner
of \fig{fig:beta}) as being similar to the MW. More
precisely, we apply the cuts $\beta>0.8$ and contribution $>50\%$, 
marked by the dashed lines. Out
of 28 Auriga galaxies, 10 ($\sim 1/3$) survive these cuts.

\textcolor{black}{We emphasize that the general conclusions from \fig{fig:beta}
do not change when we perform our analysis on all stars, 
instead of only accreted stars. We indeed tried performing the ED analysis on all 
stars, and found good fits using three components 
(1 disk component and 2 halo-like ones). We find that some of the in-situ stars 
are potentially associated with the \radial components according 
to their kinematics, and that their formation time is correlated with the 
merger time of the progenitors of these \radial components 
(more on the progenitors of the \radial components in the next section). We plan to
address the origin of such in-situ stars in a follow-up paper.}

\begin{figure}
	\resizebox{\columnwidth}{!}{\includegraphics{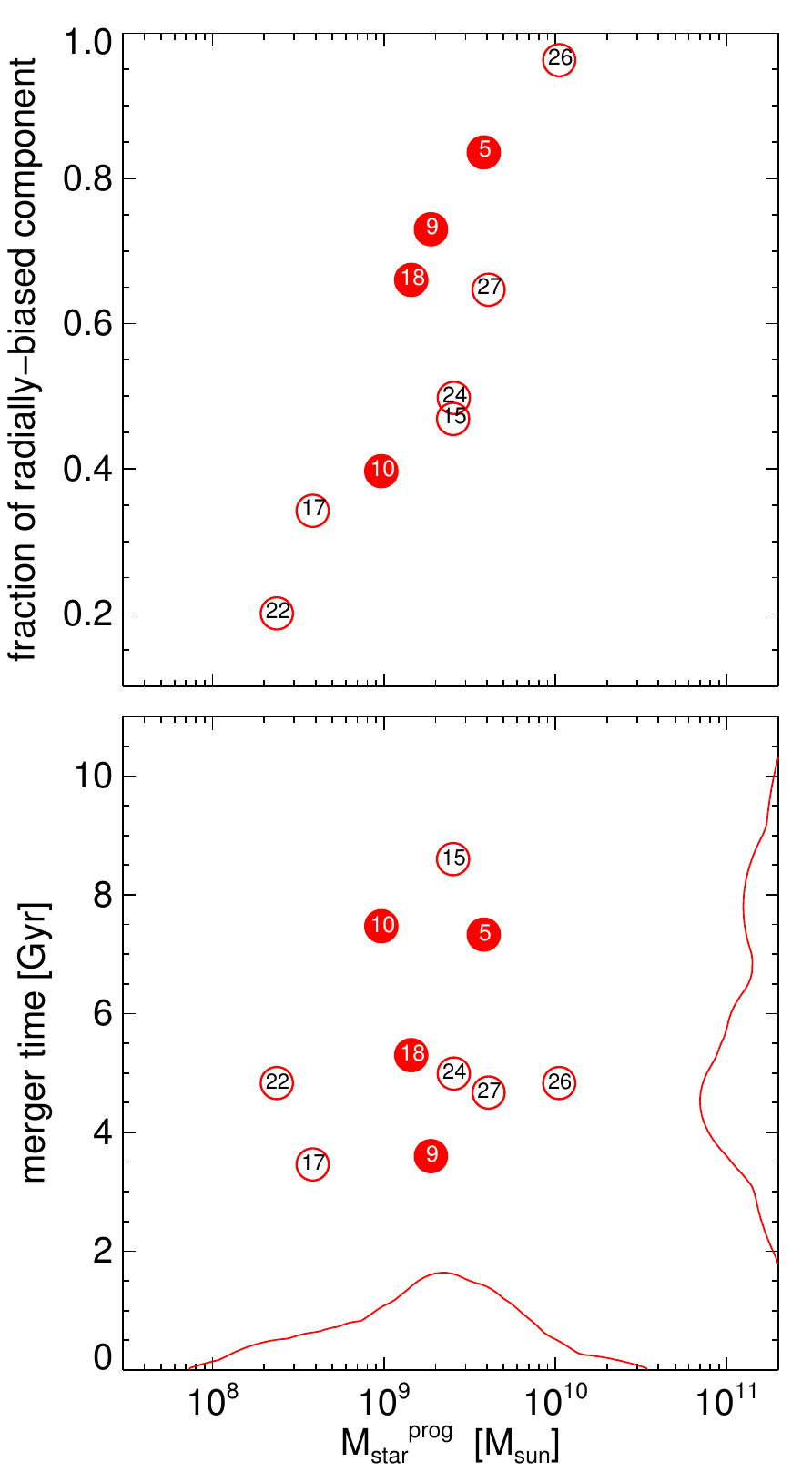}}
    \caption{{\it Top:} The stellar mass of the progenitor galaxy which 
    contributed the most to the \radial component in the Auriga galaxies, versus their mass
    contribution to that component. See text for full details of the 
    selection of the stars belonging to the \radial component. {\it Bottom:} Merger
    time of the main contributors with the [progenitors of the] MW analogues, versus 
    their peak stellar mass.
    Filled symbols correspond to the four haloes highlighted in the top-right corner of
    Fig.~\ref{fig:beta} with extremely anisotropic components dominating
    the inner halo. The distributions shown on each axis correspond to 
    projecting all of the plotted points on one axis, smoothed with a 
    Epanechnikov kernel.}
    \label{fig:prog}
\end{figure}

\subsection{Origin of the \radial component}
\label{sec:prog}

In \fig{fig:accretion} we present the dark matter accretion history of
the Auriga MW analogues. They are split into two groups, one with a significant
\radial component (red), and the rest (blue). Solid lines indicate
the median mass accreted as a function of time for each group, and the
shaded regions correspond to the full range of variation. Even though
the sample is small, it is obvious that the galaxies with a
significantly \radial component grow faster between $t=4-8
\Gyr$\footnote{$t=0 \Gyr$ corresponds to the big bang.} compared to
the rest of the galaxies. A larger sample of galaxies is, however, needed to 
confirm whether the biased assembly history is statistically significant. 
The result of \fig{fig:accretion} is broadly consistent with the proposal by
\citet{Belokurov2018} that the origin of the ``\textit{Gaia} Sausage'' is a
massive, early merger.

We investigate the origin of the prominent \radial components (i.e. $\beta>0.8$ and contribution $>50\%$) in the Auriga galaxies by tracking their star particles back in time. We select the (accreted) stars `associated' with the \radial components using: (i) metallicity, $-1.3<\feh<-0.7$; (ii) galactic height, $9<|z|/\kpc<15$; (iii) azimuthal velocity, $|V_{\theta}-V_{\rm mean}|<50\kms$; and (iv) radial velocity $\vr>100 \kms$. $V_{\rm mean}$ refers to the azimuthal velocity of the \radial component derived from the ED modeling of the previous section. The radial velocity cut is used to clean the sample from stars potentially associated with a more isotropic component. We checked that the following analysis is not sensitive to the exact values of the aforementioned cuts. \textcolor{black}{The total mass of the sample of stars `associated' with the \radial component is referred to as $M_{\rm radial}$ in the following analysis.}

\textcolor{black}{We identify all of the progenitor galaxies 
which brought in the stars that ended up in the sample of stars `associated' with the \radial component, in any target galaxy. We then rank the 
progenitors by their mass contribution ($M_{\rm c}$) to the sample.}
The progenitor that contributed the most is identified as the 
main progenitor (or main contributor) of the
\radial component\footnote{\textcolor{black}{According to our definition, the main progenitor is not necessarily the most massive progenitor, but is typically amongst the most massive ones.}}. \fig{fig:prog} presents the peak stellar mass of
the main progenitor, its relative contribution to the stars associated with the radially-biased component \textcolor{black}{(i.e. $M_{\rm c}/M_{\rm radial}$)}, 
and its merger time with the main
progenitors of the MW analogue\footnote{Merger time is defined as the
  time when {\it Subfind} fails to find the merging galaxy as a
  separate bound substructure.}.

The top panel of \fig{fig:prog} demonstrates that the stars belonging
to the \radial component predominantly come from a single progenitor: a
massive ($M_{\rm star}\sim10^9-10^{10} \Msun$) galaxy which
contributed more than $\sim50\%$ to the component. The few cases where the main
progenitor contributed less than $\sim40\%$, have one or two
additional massive progenitors. The merger time of the main
contributors, according to the bottom panel of \fig{fig:prog}, 
span the wide range $t\sim 4-8$ Gyr, with some concentration 
frequency around $t\sim4-5$ Gyr. The filled symbols, which correspond
to the four cases in the top-right corner of \fig{fig:beta}, correspond roughly to a
single LMC-like progenitor and a wide range of merger times.  The
preference for an early merger with a massive dwarf galaxy around
$t\sim4-5$ Gyr explains the behavior of the mass accretion history
(illustrated in \fig{fig:accretion}) of the galaxies with the \radial
component. The main progenitor of a \radial
component must have contributed notably to the growth of the MW
analogues.

\begin{figure}
	\resizebox{\columnwidth}{!}{\includegraphics{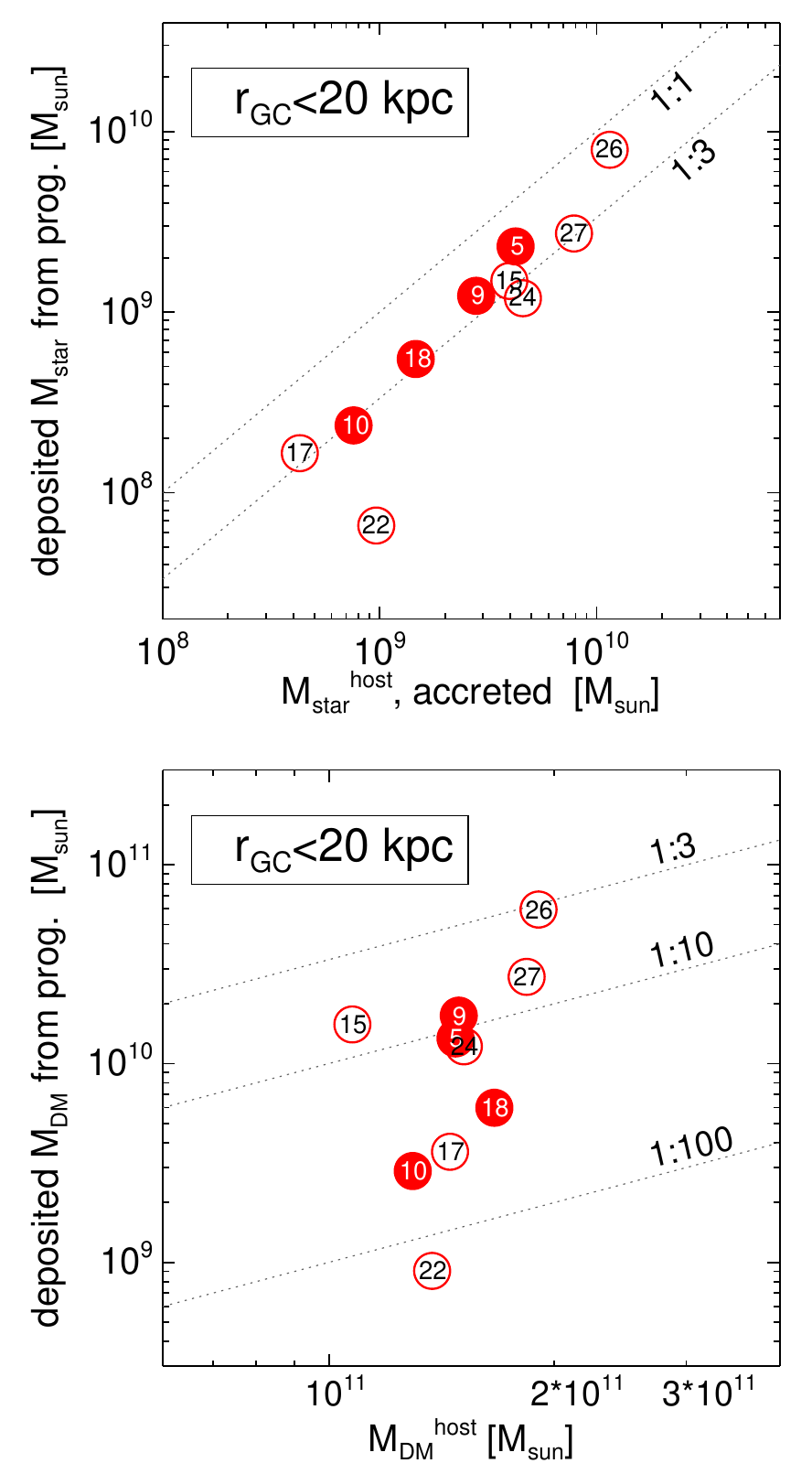}}
    \caption{{\it Top:} Contribution of the main progenitor
    of the \radial component to the stellar mass in the 
    inner 20 kpc of the Auriga galaxies at $z=0$, versus the 
    total accreted stellar mass in the same volume. 
    The diagonal lines correspond to the 1:1 and 1:3 mass ratios as indicated by the 
    labels.
    {\it Bottom:} DM contribution of the main progenitor to the inner 20 kpc, 
    versus the total DM mass in $r<20$ kpc.}
    \label{fig:dm}
\end{figure}

We now examine the contribution of the main progenitor of the \radial
component to the inner regions of the MW analogues at $z=0$. The top
panel of \fig{fig:dm} shows how much of the stellar mass in the
inner $20 \kpc$ of the MW analogues were brought in by those
progenitors, compared to the total accreted mass within $20 \kpc$. The main
progenitors of the \radial component contribute a significant
fraction to the inner stellar halo, and the contribution is
consistently $\sim1/3-1/2$ (inside $20 \kpc)$, except in one case
(Au-26). The bottom panel of \fig{fig:dm} shows that the situation is
different for the dark matter (DM): the contribution of the progenitor of the
\radial stars to the DM in the inner region ($r<20 \kpc$) varies by
more than an order of magnitude amongst different haloes, and is much
less significant compared to the stellar contribution ($\sim 1-10
\%$). There are two reasons behind the observed difference between the
DM and stellar contributions of the main progenitors. First, stars are
more centrally concentrated than the DM; hence they
occupy different regions of the energy-angular momentum space after
becoming tidally stripped. The progenitors start to
lose their DM before their stars, and therefore deposit their DM in a larger
volume. Second, the stellar mass-halo mass relation
at the faint end is so steep in $\Lambda$CDM that almost all
dwarf galaxies live in similar haloes of mass $\sim10^{10} \Msun$
\citep{Eke2004,Behroozi2013,Sawala2016a,deason16,Fattahi2018,Simpson2018}. As
a consequence, while the build up of the accreted stellar halo can be
dominated by the merger of a few relatively bright dwarf galaxies, their DM
contribution is more representative of fainter dwarfs 
\citep[see,][for a detail analysis on the build-up of MW-like haloes]{Wang2011}.

\section{Summary and Discussion}
\label{sec:discussion}

We use the Auriga magneto-hydrodynamical simulations of MW-like galaxies to
study the origin and frequency of a highly radial ($\beta\sim0.9$) and metal-rich
component  of the inner halo, recently reported by 
\citet{Belokurov2018} using \textit{Gaia}~DR1. This component is believed to be
the same population discussed by \citet{Helmi2018}.
Firstly, we present an updated version of the results of
\citet{Belokurov2018} using \textit{Gaia}~DR2. We confirm the existence of a
large group of stars on highly radial orbits with $\beta\sim0.86$ (the ``\textit{Gaia} Sausage'' ), in \textit{Gaia}~DR2.

Secondly, we search for the phase-space signatures of such a \radial population
of star particles amongst Auriga galaxies and find that they are relatively
common: around $1/3$ of the simulated galaxies have a significant population of
stars residing above and below the disc, with $\feh\sim -1$, which
resemble the observations. In addition, we find that the significance
of the sausage-like components, when they exist, always decreases at
lower $\feh$. This metallicity dependence trend has also been observed
in the MW by \citet{Belokurov2018}. Essentially, the phase space of
all galaxies looks similar at metallicities below $\feh\sim\, -2.0$
and is dominated by a halo component with $\beta\sim0.3$.

We demonstrate that the accretion history of Auriga galaxies with a \radial 
component is biased, on average, towards steeper growth at $t\sim 4-6 \Gyr$, suggesting more mergers at those early times. \textcolor{black}{It is intriguing that \citet{Mackereth2018} find, using the EAGLE simulations, a similar early accretion bias for galaxies that show alpha-bimodality in their stars. This is supported also by the study of \citet{Grand2018}, which found that the early accretion of a gas-rich major merger at around redshift 2 is required to form a prominent alpha-bimodality. The connection between alpha-bimodality and the radially-biased components is worth pursuing in follow-up studies.}

By tracing back the star particles associated with the radially-biased component of each Auriga galaxy, we find that their main progenitor is typically a massive dwarf galaxy with peak stellar mass of $10^9-10^{10} \Msun$, which merged with the progenitor of the MW analogues $\sim 6-10$ Gyr ago. Such a massive dwarf progenitor also explains why the \radial component is prominent at relatively higher metallicities ($\feh\sim\, -1$): it is a consequence of the tight stellar mass-metallicity relation \citep{Kirby2013}.   

Given that the main progenitors of the \radial components in Auriga halos are as massive as the LMC (or even more massive), their merger with the MW progenitor can have a large impact on the structure of the disc. In some cases the merger almost completely destroys the disc, and a new younger disc has to regrow. We note that these mergers are different from the those reported by \citet{Gomez2017} in the Auriga galaxies which brought in ex-situ disc stars. These have different orbits which deposit the tidally stripped stars in circular orbits co planar with the disc, as opposed to the \radial components studies here.
We will present a thorough analysis of the mergers which made the ``\textit{Gaia} Sausage''-like components, and their influence on the evolution of the MW, in a subsequent paper. 

Finally, we find that the star particles brought in by the main progenitors of
the \radial components make up $\sim 30\%-50\%$ of the total amount
of the accreted stars in the inner $20 \kpc$ of the MW analogues at
$z=0$. The dark matter contribution of the progenitors to the same
volume, however, is much lower and varies significantly between
different haloes ($\sim 1\%-10\%$). This is a consequence of the
inefficient galaxy formation at low mass scales, as well as the
difference in the tidal stripping of dark matter particles and stars
from the progenitor dwarf. Thus the accreted stellar halo
does not \emph{directly} trace the build-up of the DM halo.


\section*{Acknowledgements}

We are grateful to the referee, J. Mackereth, for his helpful comments which improved our paper. AF is supported by a European Union COFUND/Durham Junior Research
Fellowship (under EU grant agreement no. 609412). T
AD is supported by a Royal Society University Research Fellowship. AF, AD and CSF also
acknowledge the support from the STFC grant ST/P000541/1. FAG acknowledges 
financial support from FONDECYT Regular 1181264, and 
funding from the Max Planck Society through a Partner Group grant. 
RG acknowledges support by the DFG
Research Centre SFB-881 `The Milky Way System' through project A1. CSF acknowledges
support from the European Research Council (ERC) through Advanced
Investigator Grant DMIDAS (GA 786910). 
The research leading to these results has received funding from the European Research Council under the European Union's Seventh Framework Programme (FP/2007-2013) / ERC Grant Agreement n. 308024.

This work
used the DiRAC Data Centric system at Durham University, operated by
ICC on behalf of the STFC DiRAC HPC Facility (www.dirac.ac.uk). This
equipment was funded by BIS National E-infrastructure capital grant
ST/K00042X/1, STFC capital grant ST/H008519/1, and STFC DiRAC
Operations grant ST/K003267/1 and Durham University. DiRAC is part of
the National E-Infrastructure.




\bibliography{master} 



\appendix
\section{Phase-space examples of Auriga galaxies}
\label{app:example}
\begin{figure*}
	\resizebox{17.5cm}{!}{\includegraphics{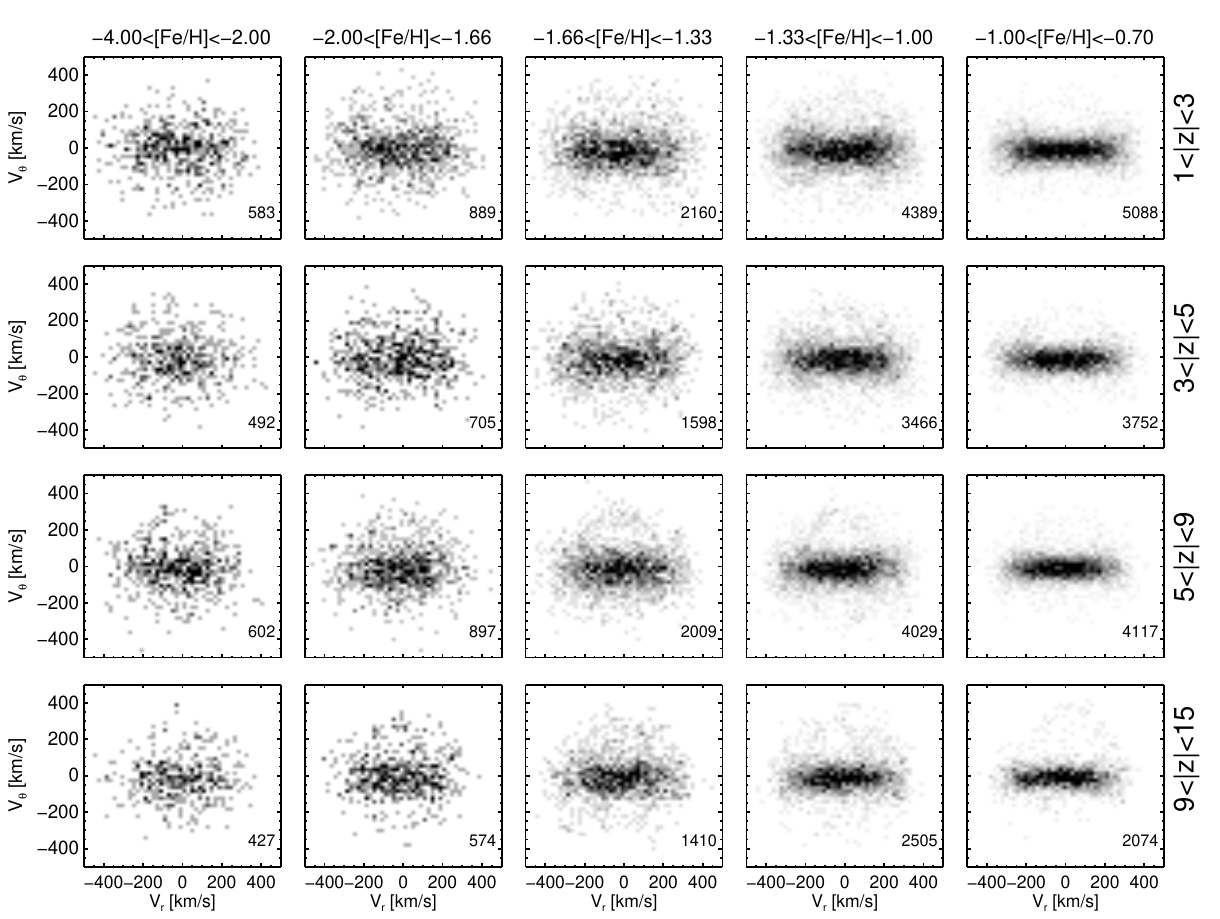}}
    \caption{ 2D velocity space (azimuthal vs radial) of accreted stars for an Auriga halo (Au-5) at different heights $|z|$ and \feh. Rows from top to bottom correspond to as $1-3$, $3-5$, $5-9$, $9-15 \kpc$ above and below the galactic disc. Different columns represent different $\feh$ of $<-2$, $-2.0,-1.7$, $-1.7,-1.3$, $-1.3, -1.0$, and $-1.0,-0.7$. This is a clear example of a galaxy with a highly radial population of stars at higher $\feh$, similar to the \textit{Gaia} observations of the MW \citep{Belokurov2018}.}
    \label{fig:Au5}
\end{figure*}

\begin{figure*}
	\includegraphics[width=17.5cm]{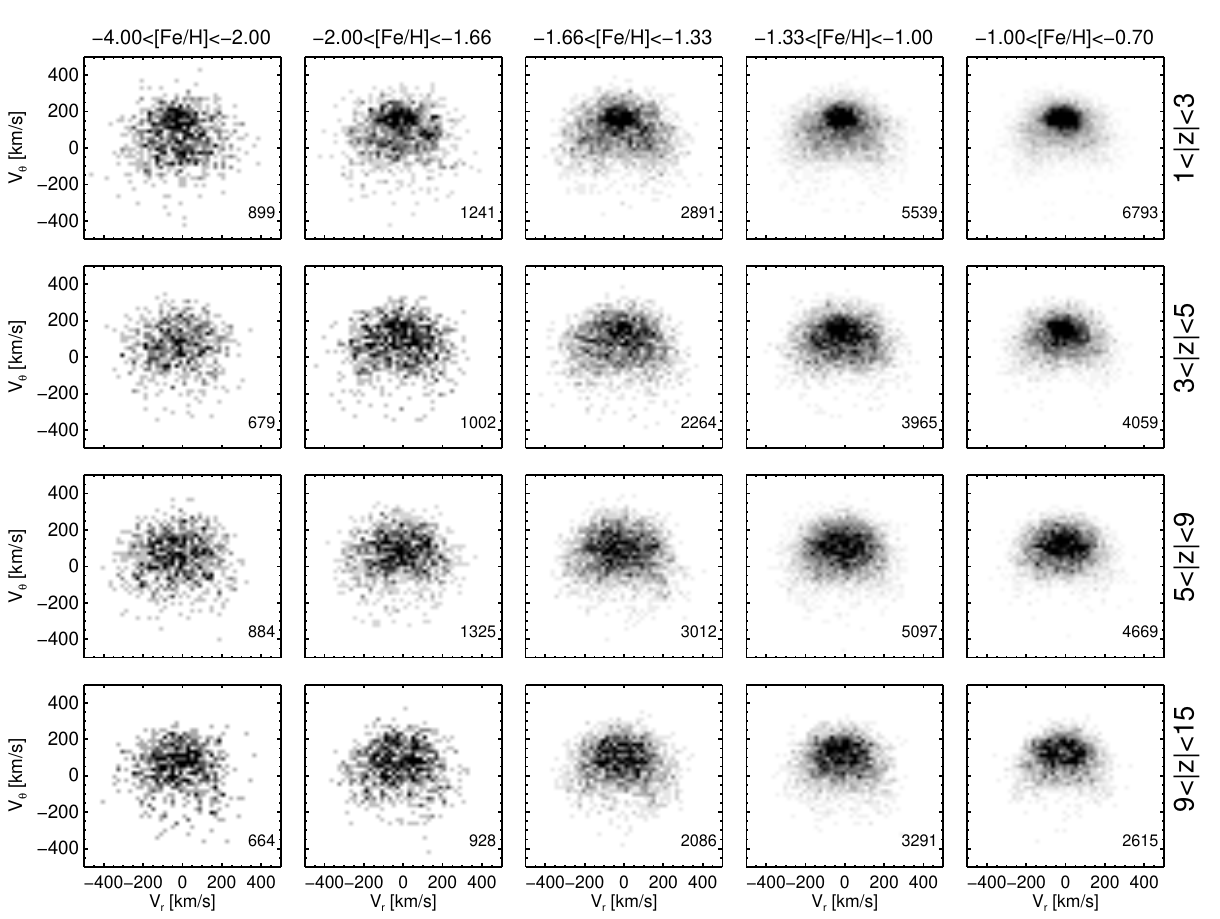}
    \caption{Similar to Fig.~\ref{fig:Au5}, but for a galaxy (Au-16) with no considerable radially-biased (``sausage''-like) component.}
    \label{fig:Au16}
\end{figure*}

Here we present the 2D velocity space of accreted stars (azimuthal vs radial components), on a full grid of height and metallicity, for 2 examples of Auriga galaxies: one where the ``\textit{Gaia} sausage''-component is prominent (\fig{fig:Au5}) and one with no notable highly radial component (\fig{fig:Au16}). The galactic height increases from the top row to bottom in the range $|z|=1-15\kpc$; $\feh$ increases from $-4$ to $-0.7$, from left to right.      

The component at higher metallicities with $v_r\sim0\kms$ and $v_{\theta}\sim200 \kms$ is the disc of the galaxies. In \fig{fig:Au5}, the highly radial component is obvious at higher metallicities, and because less and less obvious at lower metallicities. We also note that the radial component is more visible at higher $|z|$; this is due to the fact that the disc is more dominant at lower $|z|$ over the halo component. Therefore, the difference due to the existence (or lack) of the highly radial component is apparent at higher $|z|$ and higher $\feh$ (bottom-right panels).

\section{Decomposition of phase-space}
\label{app:fit}

We adopt the following, 2step procedure to decompose the phase-space of stars in each Auriga galaxy. 

\begin{figure}
	\resizebox{8cm}{!}{\includegraphics{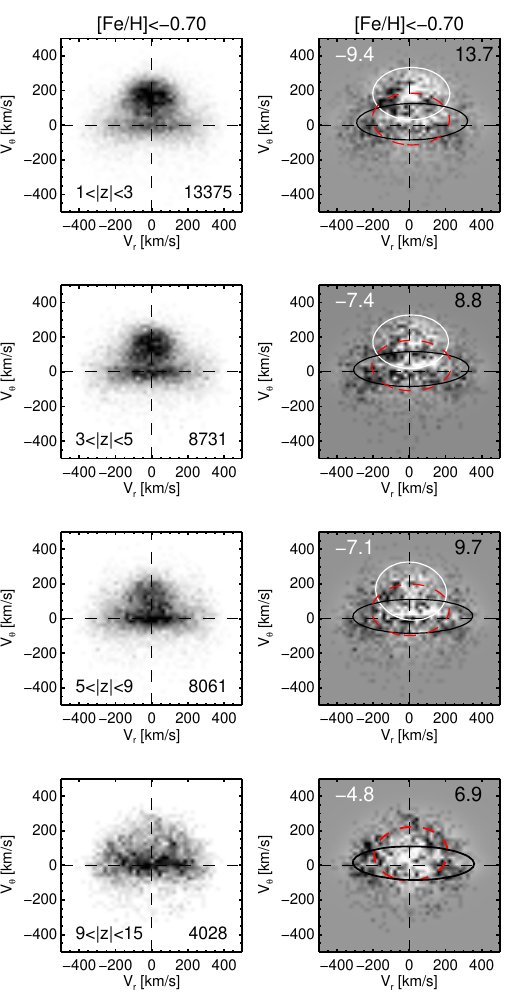}}
    \caption{An illustration of the first step of the fitting procedure: the deconvolution of the velocity space into two or three Gaussian components, after combining $\feh$ bins at each galactic height. The left column shows the velocity space at different heights (different rows), while the right column shows the residuals of the corresponding model. The three components of the model at each height include three velocity ellipsoids; one for the disc (white ellipse) and two for the halo-like components (black and red ellipses). The two component fit (last row) include only two halo-like components. The example shown is Au-6.}
    \label{fig:Au9}
\end{figure}

(i) We start with the 3D velocity space of stars with $\feh<-0.7$, at different heights. This is equivalent to stacking columns of \fig{fig:Au5}. We decompose the velocity space at each height, using the ``extreme deconvolution'' method of \citet{ED}, into three components (velocity ellipsoids) with free parameters. The initialization of parameters ensures that the three components include a disc and two halo-like components. \fig{fig:Au9} illustrates this step. 

\begin{figure*}
	\resizebox{17.5cm}{!}{\includegraphics{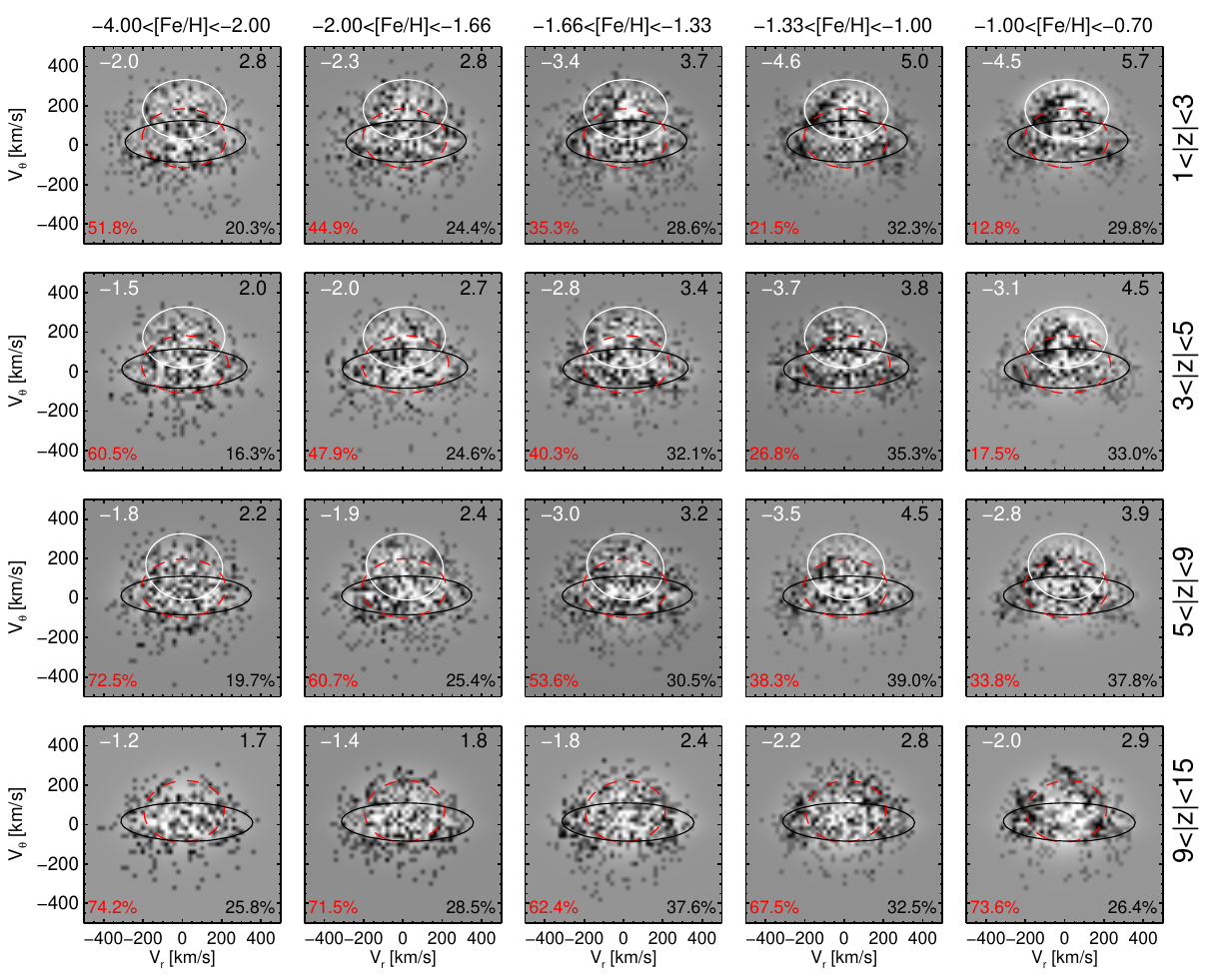}}
    \caption{Second step of the fitting procedure. The centre and shape of the velocity ellipsoids from the previous step (Fig.~\ref{fig:Au9}) are fixed, and the amplitudes of the ellipsoids vary across different $\feh$ bins. The residuals are shown here in a grid of height (rows) and metallicity (columns). The numbers in the top corners of each panel show the highest and lowest values of residuals, while the percentages in the lower corners show the contribution of different components with corresponding colors. This example is the same galaxy as \fig{fig:Au9}}
    \label{fig:Au9fit}
\end{figure*}

(ii) In the next step, we divide each height into different metallicity bins (equivalent of \fig{fig:Au5}). After fixing the centre and shape of the velocity ellipsoids from the previous step, we fit them to different metallicities and let the amplitude (contribution) vary. \fig{fig:Au9fit} shows the residuals of the fit to an example Auriga galaxy. The rows and columns are similar to Figs.~\ref{fig:Au5} and \ref{fig:Au16}. Note that the the velocity ellipses (the models) are fixed across any given row by our procedure.

\bsp	
\label{lastpage}
\end{document}